\documentclass[10pt]{article}

\begin{document}

\title{Transition from decelerated to accelerated cosmic expansion in braneworld universes}
\author{J. Ponce de Leon\thanks{E-Mail:
jponce@upracd.upr.clu.edu, jpdel@astro.uwaterloo.ca}  \\Laboratory of Theoretical Physics, 
Department of Physics\\ 
University of Puerto Rico, P.O. Box 23343,  
Rio Piedras,\\ PR 00931, USA}
\date{April 1, 2005}

\maketitle
\begin{abstract}
Braneworld theory provides a natural setting to treat, at a classical level,  the cosmological effects of  vacuum energy. Non-static extra dimensions can generally lead to a variable vacuum energy, which in turn may explain the present accelerated cosmic expansion. We concentrate our attention in models where the vacuum energy decreases as an inverse power law of the scale factor. These models agree with the observed accelerating universe, while fitting simultaneously the observational data for the density and deceleration parameter. The redshift at which the vacuum energy can start to dominate  depends on the mass density of ordinary matter. For ${\bar{\Omega}}_{m} = 0.3$, the transition from decelerated to accelerated cosmic expansion occurs at 
$z_{T} \approx 0.48 \pm 0.20$, which is compatible with SNe data. We set a lower bound on the deceleration parameter today, namely
$\bar{q} >  - 1 + 3{\bar{\Omega}}_{m}/2$, i.e., $\bar{q} > - 0.55$ for  ${\bar{\Omega}}_{m} = 0.3$. The future evolution of the universe crucially depends on the time when  vacuum starts to dominate over ordinary matter. 
If it  dominates only recently, at an epoch  $z < 0.64$, then the universe is accelerating today and will continue that way forever.  If vacuum dominates earlier, at $z > 0.64$, then the  deceleration comes back and the universe recollapses at some point  in the distant future. In the first  case, quintessence and Cardassian expansion can be formally interpreted as the low energy limit of our  model, although they are entirely different in philosophy. In the second case there is no correspondence between these models and ours.

\medskip



\end{abstract}

\newpage
\section{Introduction}
It is now widely accepted that the present universe is accelerating and spatially flat.  Evidence in favor of this is provided by observations of high-redshift supernovae Ia \cite{Riess}-\cite{Tonry}, as well as  other observations, including  the cosmic microwave background and galaxy power  spectra \cite{Lee}-\cite{Sievers}. 

Since ordinary matter is gravitationally attractive, one can conclude that the source of cosmic acceleration  is some kind of unknown matter, which violates the strong energy condition.  According to dynamical mass measurements, the total amount of ordinary matter in the universe, including dark matter,  can only account for $30\%$ of the critical density. Thus,  the other  $70\%$ corresponds to a mysterious form of matter, usually called {\it dark energy},  which remains unclustered on all scales where gravitational clustering of ordinary matter  is seen.   

Therefore, in the past few years an active field of research has been the discovery of models of the universe in which the expansion is accelerating. 
The simplest candidate for dark energy is the cosmological constant $\Lambda_{(4)}$ \cite{Peebles}-\cite{Padmanabhan0}. In this approach, $\Lambda_{(4)}$ is introduced  by ``hand"  as a parameter in Einstein's theory of
gravity.  However, if $\Lambda_{(4)}$ remains constant one faces the problem of fine-tuning or ``cosmic coincidence problem" \cite{Zlatev}, which refers to the coincidence that $\rho_{vac}$ and $\rho_{m}$ are of the same order of magnitude today.

A phenomenological solution to this problem is to consider a time dependent cosmological term, or an evolving scalar field known as {\it quintessence} \cite{Zlatev}-\cite{Deustua}. 
A variable $\Lambda_{(4)}$,  as well as quintessence,  can be modeled as the energy of a slowly evolving cosmic scalar field $\phi$ with an appropriate self-interaction potential $V(\phi)$ to account for the evolution of the universe.  The abundance  of quintessence models is due to the fact that for any given scale factor $a(t)$ and some known forms of energy density $\rho_{known}(t)$ (made of radiation, matter, etc), it is always possible to find a $V(\phi)$  that  explains the observations \cite{Padmanabhan1}.

Alternative explanations for the acceleration of the universe, beyond dark energy, include phantom energy \cite{Caldwell2}-\cite{Stefancic}, certain modifications of general relativity \cite{Turner1}-\cite{Easson}, the gravitational leakage into extra dimensions \cite{Lue}-\cite{Deffayet2}, Chaplygin gas \cite{Gorini}-\cite{Bento} as well as Cardassian models \cite{Freese}-\cite{Gondolo}. More recently, in a series of papers Vishwakarma \cite{Vishwakarma1}, \cite{Vishwakarma2} and Vishwakarma and Singh \cite{Vish and Singh}  argue that present  observations can successfully be explained  by a decelerating model of the universe in the mainstream cosmology, without invoking dark matter or dark energy.

In the present work,  we study the accelerated cosmic expansion as a consequence of embedding our observable universe as a brane in extra dimensions. Our motivation is that braneworld theory provides a natural setting to treat, at a classical level,  the cosmological effects of  vacuum energy. 
Firstly, the theory links the vacuum energy to the fundamental quantities $\Lambda_{(4)}$ and $G$. Secondly, in this theory  the vacuum energy  is completely determined by the geometry in $5D$ through Israel's boundary conditions \cite{Shiromizu}-\cite{Binetruy}. Said another way, in brane theory the cosmological term is determined by the bulk geometry and $cannot$  be put by hand as in Einstein's theory. Thirdly, non-static extra dimensions can generally lead to a variable vacuum energy \cite{jpdel1}, and consequently variable $\Lambda_{(4)}$ and $G$, which in turn may explain the present accelerated cosmic expansion.

In fact, in recent papers \cite{jpdel2}-\cite{jpdel3}  we showed that braneworld  models with variable vacuum energy (and zero or negative bulk cosmological constant) agree with the observed accelerating universe, while fitting simultaneously the observational data for the density and deceleration parameter. In those papers we were mostly interested in the general behavior of the models. We obtained  precise constrains on the cosmological parameters as well as 
demonstrated that the ``effective" equation  of state of the universe can, in principle, be determined by measurements of the deceleration parameter alone.

In this paper we are interested in more subtle details. For example, we would like 
to know the redshift of transition from deceleration to acceleration. Namely, based on our model, can we predict, or at least narrow down, the redshift interval for the transition?  What is the time at which the vacuum energy should  start to dominate in order to explain the observed cosmic acceleration?  How is it related to the current   mass density of the universe? Does the  present deceleration parameter depend on it? What is the future evolution of the universe? Will it continue to accelerate forever? 

In the scenario discussed here, the acceleration of the universe is related not only to the variation of vacuum energy and cosmological term, but also to the time evolution of $G$ and, possibly, to the variation of other fundamental ``constants" as well. 

The paper is organized as follows. In section $2$ we give a brief summary of the equations for homogeneous cosmologies in $5D$ based on braneworld theory. In section $3$ we show how to incorporate a varying vacuum energy into the scheme. Observational constraints on the model are discussed in section $4$. In section $5$ we discuss the transition from deceleration to acceleration. 

We show that the crossover point strongly depends on the energy density of ordinary matter. The possible values of the redshift of transition are spread over a finite interval, regardless of the moment in time when  vacuum starts to dominate over ordinary matter. This is a clear indication that the effects of vacuum rapidly decrease with the increase of $z$.
We also  show that the value of the deceleration parameter today is bounded bellow. 

We discuss the cosmological parameters for the case where the matter density today is one-third of the critical density. We find  that the future evolution of the universe crucially depends on the time when  vacuum starts to dominate over ordinary matter. Namely, if the vacuum contribution starts to  dominate only recently, at an epoch  $z < 0.64$, then the universe is accelerating today and will continue that way forever.  But, if the vacuum  dominates earlier, at $z > 0.64$, then the  deceleration comes back and the universe recollapses at some point  in the distant future.

Finally, in section $6$ we give a summary and discussion. We show that the original Cardassian expansion, proposed by Freese and Lewis \cite{Freese}, as well as dark energy models (with constant $w_{X}$) can be interpreted as the low energy limit of our ever-expanding models.

\section{Homogeneous cosmology in $5D$}
In order to facilitate the discussion, and set the notation, we start with a brief summary of the pertinent ideas and equations in the braneworld scenario. In this scenario our homogeneous and isotropic universe is envisioned as a singular hypersurface embedded in a five-dimensional  manifold with metric 

\begin{equation}
\label{cosmological metric}
d{\cal{S}}^2 = n^2(t,y)dt^2 - a^2(t,y)\left[\frac{dr^2}{(1 - kr^2)} + r^2(d\theta^2 + \sin^2\theta d\phi^2)\right] - \Phi^2(t, y)dy^2,
\end{equation}
where $t, r, \theta$ and $\phi$ are the usual coordinates for a spacetime with spherically symmetric spatial sections and $k = 0, +1, -1$.  The metric is a solution of the five-dimensional Einstein equations
\begin{equation}
\label{field equations in 5D}
{^{(5)}G}_{AB} = {^{(5)}R}_{AB} - \frac{1}{2} g_{AB}{^{(5)}R} = {k_{(5)}^2} {^{(5)}T_{AB}}, 
\end{equation}
where ${^{(5)}T}_{AB}$ is the five-dimensional energy-momentum tensor and $k_{(5)}$ is a constant introduced for dimensional considerations. 
 
The energy-momentum tensor on the brane $\tau_{\mu\nu}$ is separated in  two parts, 
\begin{equation}
\label{decomposition of tau}
\tau_{\mu\nu} =  \sigma g_{\mu\nu} + T_{\mu\nu},
\end{equation} 
where $\sigma$ is the tension of the brane in  $5D$, which is interpreted as the vacuum energy of the braneworld, and $T_{\mu\nu}$ represents the energy-momentum tensor of ordinary matter in $4D$. 

 There are two assumptions relating the physics in $4D$ to the geometry of the bulk. They are:
(i) the  bulk spacetime possesses   ${\bf Z_{2}}$ symmetry about the brane, and  
(ii) the brane is embedded in an Anti-de Sitter bulk, i.e., ${^{(5)}T}_{AB}$ is taken as 
\begin{equation}
\label{AdS}
{^{(5)}T}_{AB} =  \Lambda_{(5)}g_{AB}, 
\end{equation}
where $\Lambda_{(5)} < 0$.

As a consequence of the first assumption, the matter and vacuum energy density in $4D$ become completely  determined by the geometry in $5D$ through Israel's boundary conditions. Namely, for a perfect fluid 
\begin{equation}
\label{EMT for perfect fluid}
T_{\mu\nu} = (\rho + p)u_{\mu}u_{\nu} - p g_{\mu\nu},
\end{equation}
with 
\begin{equation}
\label{equation of state for ordinary matter}
p = \gamma \rho, 
\end{equation}
the matter density is given by \cite{jpdel1}
\begin{equation}
\label{density}
\rho = \frac{(- 2)}{k_{(5)}^2 (\gamma + 1)\Phi|_{brane}}\left[\frac{a'}{a} - \frac{n'}{n}\right]_{brane}.
\end{equation}
and the vacuum density is
\begin{equation}
\label{tension of the brane}
\sigma =   \frac{(- 2)}{k_{(5)}^2 (\gamma + 1)\Phi|_{brane}}\left[(3\gamma + 2)\frac{a'}{a} + \frac{n'}{n} \right]_{brane},
\end{equation}
where a prime denotes derivative with respect to the extra variable $y$. 

 The second assumption has two important consequences. Firstly, since ${^{(5)}T}_{\mu 4} = 0$, it follows that the energy-momentum tensor on the brane $\tau_{\mu\nu}$ is a conserved quantity, viz.,
\begin{equation}
\label{conservation on the brane}
\tau^{\nu}_{\mu; \nu} = 0. 
\end{equation}
Secondly, the field equations (\ref{field equations in 5D}) admit a first integral, namely, 
\begin{equation}
\label{first integral in the bulk}
\left(\frac{\dot{a}}{an}\right)^2 =  \frac{k_{(5)}^2 \Lambda_{(5)}}{6} + \left(\frac{a'}{a \Phi}\right)^2 - \frac{k}{a^2} + \frac{\cal{C}}{a^4}, 
\end{equation}
where ${\cal{C}}$ is  a constant of integration which arises from the projection of the Weyl curvature tensor of the bulk on the brane.   
Evaluating (\ref{first integral in the bulk}) at the brane, which is fixed at some $y = y_{brane} = const$, as well as  using (\ref{density}) and (\ref{tension of the brane}), we obtain the generalized Friedmann equation, viz., 
\begin{equation}
\label{generalized FRLW equation}
3\left(\frac{{\dot{a}}_{0}}{a_{0}}\right)^2  = \Lambda_{(4)}  + 8 \pi G \rho + \frac{ k_{(5)}^4}{12}\rho^2 - \frac{3 k}{a_{0}^2} + \frac{3 {\cal{C}}}{a_{0}^{4}}, 
\end{equation}
where $a_{0}(t) = a(t, y_{brane})$,  and  
\begin{equation}
\label{definition of lambda}
\Lambda_{(4)} = \frac{1}{2}k_{(5)}^2\left(\Lambda_{(5)} + \frac{ k_{(5)}^2 \sigma^2}{6}\right),
\end{equation}
\begin{equation}
\label{effective gravitational coupling}
8 \pi G =  \frac{k_{(5)}^4 \sigma}{6}.
\end{equation}
These quantities are interpreted as the  {\em net} cosmological term and gravitational coupling in $4$ dimensions, respectively. 

Equation (\ref{generalized FRLW equation}) contains two novel features; it relates  the   fundamental quantities $\Lambda_{(4)}$ and $G$ to the vacuum energy, and carries  higher-dimensional modifications to the Friedmann-Robertson-Walker (FRW) cosmological models of general relativity. Namely, local quadratic corrections via $\rho^2$, and the nonlocal corrections from the free gravitational field in the bulk, transmitted  by the dark radiation term ${\cal{C}}/{a^4}$. 

Except for the condition that $n = 1$ at the brane, the generalized Friedmann  equation (\ref{generalized FRLW equation}) is valid for {\em arbitrary} $\Phi(t,y)$ and $n(t,y)$ in the bulk \cite{Binetruy}. This equation allows us to  examine the evolution of the brane without using any particular solution of the five-dimensional field equations. In what follows we will omit the subscript $0$.

\section{Variable vacuum energy}

In equation (\ref{generalized FRLW equation}), $G$ and $\Lambda_{(4)}$ are usually assumed to be ``truly" constants. However, the vacuum energy density $\sigma$ does not have to be a constant. From (\ref{tension of the brane}) it follows that $\sigma$ depends on the details of the model. Indeed, we have recently shown \cite{jpdel1} that there are several models, with reasonable physical properties, for which a variable $\Phi$ induces a variation in the vacuum energy $\sigma$. 

\subsection{Variable vacuum: an example}
As an illustration, let us consider the class of solutions to the field equations (\ref{field equations in 5D}) for which the metric coefficients in (\ref {cosmological metric}) are separable functions of their arguments. For this class of solutions, without loss of generality we can set
\begin{equation}
n = n(y), \;\;\; a(t,y) = \tilde{a}(t)Y(y), \;\;\;\; \Phi = \Phi(t).
\end{equation}
From $G_{04} = 0$ it follows that 
\begin{equation}
\label{separation of variables}
\left(\frac{n'}{n}\right) = \zeta \left(\frac{Y'}{Y}\right),\;\;\;\frac{\dot{\Phi}}{\Phi} = (1 - \zeta)\frac{\dot{\tilde{a}}}{\tilde{a}},
\end{equation}
where $\zeta$ is a separation constant. Consequently, 
 \begin{equation}
\label{solution for Phi in the case of separation of variables}
\Phi(t) = A {\tilde{a}}^{(1 - \zeta)},
\end{equation}
 where $A$ is a constant of integration. Thus, for any $\zeta \neq 1$, $\Phi$ is a variable function of $t$. In what follows, for simplicity of the notation,    instead of $\zeta$  we will use $\beta = - (\zeta + 2)/3$ (or $\zeta  = - 3 \beta - 2$). With this notation we have $\Phi(t) = A {\tilde{a}}^{3(\beta + 1)}$. Substituting this expression into (\ref{tension of the brane}) we find
\begin{equation}
\label{vacuum energy from the brane}
\sigma = \frac{D (\gamma - \beta)}{(\beta + 1)a ^{3(\beta + 1)}},
\end{equation}
where we have introduced the constant $D \equiv - 6 (\beta + 1)Y'_{brane}Y_{brane}^{3\beta + 2}/A(\gamma + 1)k_{(5)}^2$.  This epitomizes the general situation where $\sigma$ is a function of time. It is worth noting that only the assumption of separability of the bulk metric (\ref{cosmological metric}) underlies equation (\ref{vacuum energy from the brane}), i.e., we do not need to know the details  of the solution in the five-dimensional bulk.

From a physical point of view,  the vacuum energy (\ref{vacuum energy from the brane}) implies $(\dot{G}/G) = - 3(\beta + 1)H$. Conversely, if we extrapolate the present limit $|\dot{G}/G| = |g| H$, and assume that $g$ is a {\it constant}\footnote{The physical meaning of this assumption 
is that the  variation of $g$ is much ``slower" than that of $H$ and $G$. Namely, $|\dot{g}/g| << |\dot{H}/H|$, $|\dot{g}/g| << |\dot{G}/G|$},  then we obtain a cosmological model where the vacuum energy  is given exactly by (\ref{vacuum energy from the brane}),  with $g = - 3(\beta + 1)$. This model is consistent with the observed acceleration and flatness of our universe \cite{jpdel2}. 
 
\subsection{Our model}
In general, for variable vacuum energy, the conservation equations (\ref{conservation on the brane}) for a perfect fluid which satisfies (\ref{equation of state for ordinary matter}), yield 
\begin {equation}
\label{conservation equation in explicit form}
\dot{\rho} + 3\rho(\gamma + 1)\frac{\dot{a}}{a} =  - \dot{\sigma},
\end{equation}
From which it follows that
\begin{equation}
\label{general solution for rho}
\rho = \frac{C}{a^{3(\gamma + 1)}} - \frac{1}{a^{3(\gamma + 1)}}\int{a^{3(\gamma + 1)}d\sigma},
\end{equation}
where $C$ is a constant of integration. For the case of {\em constant} $\sigma$, we recover  the familiar relationship between the matter energy  density and the expansion factor $a$, viz., 
\begin{equation}
\label{familiar expansion factor}
\rho =  \frac{C}{a^{3(\gamma + 1)}}.
\end{equation}
 The second term in (\ref{general solution for rho}) is the contribution associated with the variation of vacuum. {The variation of the vacuum energy is deeply rooted in fundamental physics. The simplest microphysical model for a variable $\sigma$,  as well as for $\Lambda_{(4)}$ and  quintessence,  is the energy associated with a slowly evolving cosmic scalar field $\phi$ with some self-interaction potential $V(\phi)$ minimally coupled to gravity. The  
potentials  are suggested by particle physics, but in principle  $V(\phi)$ can be determined from supernova observations \cite{Huterer}-\cite{Gerke}.}  

In this paper,  instead of constructing a field theory for the time evolution of the vacuum energy\footnote{According to Padmanabhan \cite{Padmanabhan1} it is trivial to choose the ``appropriate" potential $V (\phi)$ such that we can explain the observations, 
for any given pair $a(t)$ and  $\rho(t)$}, we employ  our previous example  as a guide. Namely, if during the expansion of the universe $\sigma$ decays as in  (\ref{vacuum energy from the brane}), then  from (\ref{general solution for rho}) it follows that
\begin{equation}
\label{assumption for the density}
\rho = \frac{C}{a^{3(\gamma + 1)}} + \frac{D}{a^{3(\beta + 1)}}.  
\end{equation}
Now,  from  the  conservation  equation  (\ref{conservation equation in explicit form})  we get 
\begin{equation}
\label{variable density}
\sigma = \sigma_{0} + \frac{D(\gamma - \beta)}{(\beta + 1)a^{3(\beta + 1)}},
\end{equation}
where $\sigma_{0}$ is a constant of integration. We will assume this form, with $\sigma_{0} \neq 0$,   for the vacuum energy\footnote{With this assumption $g$ is not constant.}. It  reduces to the case of constant vacuum energy  $\sigma = \sigma_{0}$,  for $\beta = \gamma$.  However,  for $\beta \neq \gamma$ it  generates models with variable $\sigma$, $G$ and $\Lambda_{(4)}$.   
We immediately notice that the positivity of $G = k_{(5)}^4 \sigma/6$ requires $\beta < \gamma$. 

\subsubsection{The effective density}

The total energy density of the universe, $\rho_{total} = \sigma + \rho$, can be written as 
\begin{equation}
\label{effective density}
\rho_{total} = \sigma_{0} + \rho_{eff},
\end{equation}
where we have introduced the ``effective" density $\rho_{eff}$ as 
\begin{equation}
\label{effective density}
\rho_{eff} = \frac{C}{a^{3(\gamma + 1)}}  + \left(\frac{\gamma + 1}{\beta + 1}\right)\frac{D}{a^{3(\beta + 1)}}.
\end{equation}
This effective density is the one that drives the evolution of the universe. Indeed, in the present model  the generalized Friedmann equation becomes 
\begin{equation}
\label{gen. F. eq. for the case under study}
3\left(\frac{\dot{a}}{a}\right)^2 = \frac{1}{2}k_{(5)}^2 \left(\Lambda_{(5)} + \frac{k_{(5)}^2}{6}\sigma_{0}^2\right) + 
\frac{k_{(5)}^4 \sigma_{0}}{6}\rho_{eff}
 +  \frac{k_{(5)}^4 }{12}\rho_{eff}^2 - \frac{3 k}{a^2} + \frac{3 {\cal{C}}}{a^{4}}. 
\end{equation}
We note that, distinct from (\ref{assumption for the density}), the vacuum contribution   in the effective density (\ref{effective density}) is multiplied by the factor $(\gamma + 1)/(\beta + 1)$, which is larger than $1$, because $\beta < \gamma$. 

In what follows we will set $\gamma = 0$, in view of the fact that our universe is matter-dominated $(p = 0)$. In addition, we can set $k =  0$, because astrophysical data from BOOMERANG \cite {BOOMERANG} and WMAP \cite{Spergel} indicate that our universe is  flat. Also we can consider ${\cal{C}} = 0$, since the constant $\cal{C}$, which  is an effective radiation term related to the bulk Weyl tensor, is constrained to be small enough at the time of 
nucleosynthesis and it should be negligible today. Also, in order to avoid an exponential expansion  of the universe in its asymptotic limit, we assume
\begin{equation}
\label{avoiding exp. and rec.}
\Lambda_{(5)} + \frac{k_{(5)}^2}{6}\sigma_{0}^2 = 0.
\end{equation}
Thus, in the case under consideration the evolution of the universe  will be governed by
\begin{equation}
\label{F. eq. for the case under study}
3\left(\frac{\dot{a}}{a}\right)^2 =  
\frac{k_{(5)}^4 \sigma_{0}}{6}\rho_{eff} + \frac{k_{(5)}^4}{12}\rho_{eff}^2,
\end{equation}
with
\begin{equation}
\label{eff. density for dust}
\rho_{eff} = \frac{C}{a^3}  + \frac{D/(\beta + 1)}{a^{3(\beta + 1)}}. 
\end{equation}
The cosmological term $\Lambda_{(4)}$ is not constant, but  evolves according to 
\begin{equation}
\label{cosmological constant in terms of a}
\Lambda_{(4)} = \frac{k_{(5)}^4}{6}\frac{\sigma_{0}D(- \beta)}{(\beta + 1)a^{3(\beta + 1)}} + \frac{k_{(5)}^4}{12}\frac{D^2\beta^2}{(\beta + 1)^2a^{6(\beta + 1)}}.
\end{equation} 

\subsubsection{Asymptotic behavior}

For $\beta = \gamma = 0$, we recover the usual picture, i.e., $\rho = \bar{\rho}(1 + z)^{3}$,  $\sigma = \sigma_{0}$ and $\Lambda_{(4)} = 0$. For  $\beta < \gamma  = 0$, the vacuum term is initially negligible, which means that $\rho_{eff}$  approaches  the typical matter density in FRW models and $(\sigma/\rho) \rightarrow 0$.

If $\sigma_{0}$ is positive, then the  universe is in continuous expansion. When the vacuum term  in (\ref{eff. density for dust}) is so large that the ordinary matter contribution can be neglected, we find  
\begin{equation}
\label{evolution for small z}
a^{3(\beta + 1)} \approx \frac{k_{(5)}^4\sigma_{0}D (\beta + 1)}{8}t^2. 
 \end{equation}
The corresponding ``deceleration" parameter $q = - \ddot{a}a/{\dot{a}}^2$ is given by 
\begin{equation}
\label{deceleration 4.1}
q \approx \frac{1 + 3 \beta}{2},
\end{equation}
which indicates that the expansion is accelerated for $\beta < - 1/3$. At this stage of the evolution $G$ is constant and the  cosmological term $\Lambda_{(4)}$ varies with time. Namely,
\begin{equation}
\label{effective Lambda 4.1}
8\pi G \approx k_{(5)}^4 \frac{\sigma_{0}}{6}, \;\;\;  \Lambda_{(4)} \approx \frac{3(\gamma - \beta)}{(\gamma + 1)}H^2, \;\;\; H \approx \frac{2}{3(\beta + 1)} \frac{1}{t} .
\end{equation}

\subsubsection{The vacuum energy takes over}

We note that, for $\beta < 0$, the first term in (\ref{eff. density for dust}) decreases in time faster than the second one. Therefore, $D$ has to be  chosen in such a way as to make the second term in  (\ref{eff. density for dust}) important at the right time to explain the observations. In order to do this, 
we find useful to introduce the  {\it auxiliary}  quantity $z_{eq}$. This is  the redshift at which the vacuum and matter components in (\ref{eff. density for dust}) become equal to each other\footnote{In order to avoid misunderstanding: the parameter $z_{eq}$ is not the redshift of transition from deceleration to acceleration, which we  denote as $z_{T}$.}, i.e., $C/a^{3} = D/(\beta + 1)a^{3(\beta + 1)}$ at $a = a(z_{eq})$. Since $a = \bar{a}/(1 + z)$, where $\bar{a}$ is the present value of $a$, we obtain

\begin{equation}
\label{C in terms of D}
C = D \frac{(1 + z_{eq})^{3\beta }}{(\beta + 1){\bar{a}}^{3\beta}}.
\end {equation}
The appropriate $D$ follows from the evaluation of $\rho_{eff}$ today.   We find
\begin{equation}
\label{D in terms of present rho}
\frac{D}{{\bar{a}}^{3(\beta + 1)}} = \frac{(\beta + 1){\bar{\rho}}_{eff}}{[1 + (1 + z_{eq})^{3\beta}]}, 
\end{equation}
where ${\bar{\rho}}_{eff}$ is the present value of the effective density.

We note that the case where $D = 0$ and $C \neq 0$ is {\it formally} obtained from our equations by setting $z_{eq} = - 1$, for any $\beta < 0$.  Similarly, the case where $C = 0$ and $D \neq 0$ is formally attained in the limit $z_{eq} \rightarrow \infty$.

\section{Observational constraints on $\beta$}

Although the evolution equations (\ref{F. eq. for the case under study})-(\ref{eff. density for dust}) contain four constants: $C$, $D$, $\sigma_{0}$, and $k_{(5)}^4$,  there are only two parameters in the model, viz.,  $\beta$ and $z_{eq}$.
The aim of this section is to find out the physical restrictions on the parameter $\beta$. 

With this goal, we express the relevant quantities in terms of these parameters and the present-value of the density of ordinary matter $\rho_{m} = C/a^3$.
Let us start with the effective density (\ref{eff. density for dust}). Using (\ref{C in terms of D}) and (\ref{D in terms of present rho}),  it can be written as
\begin{equation}
\label{effective density in terms of matter density}
\rho_{eff} = {\bar{\rho}}_{eff}\frac{(1 + z_{eq})^{3\beta}}{[1 + (1 + z_{eq})^{3\beta}]}(1 + z)^3\left[1 + \left(\frac{1 + z}{1 + z_{eq}}\right)^{3\beta}\right].
\end{equation}
The ratio of effective density to ordinary matter density is given by
\begin{equation}
\label{ratio of eff density to matter density}
\frac{\rho_{eff}}{\rho_{m}} = 1 + \left(\frac{1 + z}{1 + z_{eq}}\right)^{3\beta}.
\end{equation}
Thus, at the present time
\begin{equation}
\label{rho eff to rho matter at the present}
{\bar{\rho}}_{eff} = {\bar{\rho}}_{m}F(z_{eq}, \beta),
\end{equation}
where
\begin{equation}
\label{def. of F}
F(z_{eq}, \beta) = \frac{[1 + (1 + z_{eq})^{3\beta}]}{(1 + z_{eq})^{3\beta}}.
\end{equation}
With this notation, we have $C = {\bar{a}}^3{\bar{\rho}}_{m}$, $D = {\bar{a}}^{3(\beta + 1)}(F - 1)(\beta + 1){\bar{\rho}}_{m}$. Consequently,  

\begin{equation}
\label{eff. density in terms of F}
\rho_{eff} = \rho_{m}[1 + (F - 1)(1 + z)^{3\beta}].
\end{equation}

Thus,  in our formulae the case of constant vacuum energy, for which $D = 0$ and $\rho_{eff} = \rho_{m}$, corresponds to $F =  1$, for any $\beta < 0$. For large redshifts $\rho_{eff} \approx \rho_{m}$, while at the present time ${\bar{\rho}}_{eff} = {\bar{\rho}}_{m}F$. We note that $F$ can be very large for large values of the parameter $Z_{eq}$. 

\subsection{Positivity of $G$}

We now proceed to calculate  $\sigma_{0}$. Evaluating (\ref{F. eq. for the case under study}) today, and using (\ref{rho eff to rho matter at the present}), we have
\begin{equation}
\label{eval. of H}
3{\bar{H}}^2 = \frac{k_{(5)}^4}{6}\left[\sigma_{0}{\bar{\rho}}_{m}F(z_{eq}, \beta) + \frac{1}{2}{\bar{\rho}}_{m}^2F^2(z_{eq}, \beta)\right].
\end{equation}
The constant $k_{(5)}^4$ is given by $k_{(5)}^4 = 48 \pi G/\sigma$. Thus, using (\ref{variable density}), with $\gamma = 0$,  (\ref{D in terms of present rho}) and (\ref{def. of F}), we obtain
\begin{equation}
\label{k five}
k_{(5)}^4 = \frac{48 \pi \bar{G}}{[\sigma_{0} - \beta {\bar{\rho}}_{m}(F - 1)]},
\end{equation}
where $\bar{G}$ is the present value of the gravitational ``constant" $G$. 
Feeding this expression back into (\ref{eval. of H}) we obtain
\begin{equation}
\label{sigma 0 as a function of the parameters and rho bar}
\sigma_{0} = {\bar{\rho}}_{m}\frac{[{\bar{\Omega}}_{m}F^2/2 + \beta(F - 1)]}{[1  - {\bar{\Omega}}_{m}F]},
\end{equation}
where ${\bar{\Omega}}_{m}$ is the present value of the mater density parameter $\Omega_{m} = 8 \pi G \rho_{m}/3H^2$. The vacuum term (\ref{variable density}) can be written as
\begin{equation}
\label{sigma as a function of z}
\sigma(z, z_{eq}, \beta) = \sigma_{0} -\beta {\bar{\rho}}_{m}(F - 1)(1 + z)^{3(\beta + 1)}.
\end{equation}
Evaluating this equation today, we have
\begin{equation}
\label{total sigma}
\bar{\sigma}(z_{eq}, \beta, {\bar{\Omega}}_{m}) = {\bar{\rho}}_{m} \frac{{\bar{\Omega}}_{m}F[F/2 + \beta(F - 1)]}{1 - {\bar{\Omega}}_{m}F}
\end{equation}

We now calculate  the constant $k_{(5)}^4$.  For this we substitute (\ref{sigma 0 as a function of the parameters and rho bar}) into (\ref{k five}). We get
\begin{equation}
\label{k}
k_{(5)}^4(z_{eq}, \beta) = \frac{18 {\bar{H}}^2}{{\bar{\rho}}_{m}^2}\frac{[1 - {\bar{\Omega}}_{m}F]}{[F/2 + \beta(F - 1)]F}.
\end{equation}
We are now able to express general physical  conditions on $\beta$. 
Since $8\pi {G} = k_{(5)}^4 {\sigma}/6$, the positivity of $G$ demands $k_{(5)}^4 {\sigma} > 0$. Thus, for any given ${\bar{\Omega}}_{m}$ and $z_{eq}$, the allowed values of $\beta $, in (\ref{total sigma})  and (\ref{k}), are those for which\footnote{We note that $\sigma \geq \bar{\sigma}$. Also, the possibility 
$k_{(5)}^4(z_{eq}, \beta) <  0$, and $\bar{\sigma}(z_{eq}, \beta, {\bar{\Omega}}_{m}) < 0$ is excluded by the fact that $k_{(5)}^2$ in (\ref{field equations in 5D}) is a real number.} 
\begin{equation}
\label{theoretical conditions on beta}
k_{(5)}^4(z_{eq}, \beta, {\bar{\Omega}}_{m}) > 0,\;\;\;\; \bar{\sigma}(z_{eq}, \beta, {\bar{\Omega}}_{m}) > 0.
\end{equation}

\subsection{$\dot{G}/G$}

More stringent  restrictions on $\beta$ follow from observational constraints on the variation of $G$. In terms of the Hubble parameter, the time evolution of $G$  is usually written as  $(\dot{G}/G) = g H$, where $g$ is a 
dimensionless parameter. In our model we have 
\begin{equation}
\label{Gdot over G}
g = - 3(\beta + 1)\left(1  - \frac{\sigma_{0}}{\sigma}\right).
\end{equation}
We note that $g = 0$ for $\beta = \gamma$ (i.e.,  $\beta = 0$ for ordinary matter). From (\ref{sigma 0 as a function of the parameters and rho bar}) and (\ref{total sigma}), we obtain the present value of $g$ as
\begin{equation}
\label{g}
\bar{g}(z_{eq}, \beta, {\bar{\Omega}}_{m}) = - \frac{3\beta(\beta + 1)(F - 1)({\bar{\Omega}}_{m}F - 1)}{{\bar{\Omega}}_{m}F[F/2 + \beta(F - 1)]}
\end{equation}
The abundance of various elements as well as  
nucleosynthesis are used to put 
constraints on $g$ today. The present observational upper 
bound is\footnote{A comprehensive and updated discussion  of the various experimental and 
observational constraints on the value of $g$ (as well as on the variation 
of other fundamental ``constants" of nature) has recently been provided 
by Uzan \cite{Uzan}} 

\begin{equation}
\label{upper bound for g}
|\bar{g}| \leq  0.1.
\end{equation}
Thus, the choice of $\beta$ has to guarantee the fulfillment of this condition. It clearly pushes the values of $\beta$ either toward $\beta \approx 0$ or $\beta \approx - 1$.

\subsection{Deceleration parameter $q$}

In order to consider other observational constraints on $\beta$, let us  introduce the deceleration parameter

\begin{equation}
\label{q in terms of z}
q = - \frac{\ddot{a} a}{{\dot{a}}^2} = - 1 + \frac{(1 + z)}{H}\frac{dH}{dz},
\end{equation}
which in the case under study becomes

\begin{equation}
\label{acceleration}
q(z) = - 1 + \frac{3(\sigma_{0} + \rho_{eff})}{(2 \sigma_{0} + \rho_{eff})}\frac{[1 + (\beta + 1)(F -1)(1 + z)^{3\beta}]}{[1 + (F - 1)(1 + z)^{3\beta}]},
\end{equation}
where $\rho_{eff}(z)$ and $\sigma_{0}$ are given by (\ref{ratio of eff density to matter density}) and (\ref{sigma 0 as a function of the parameters and rho bar}), respectively.
Evaluating this equation today, we have

\begin{equation}
\label{present day acceleration}
\bar{q}(z_{eq}, \beta, {\bar{\Omega}}_{m}) = - 1 + 3\frac{[F + \beta(F - 1) - {\bar{\Omega}}_{m}F^2/2][F + \beta(F - 1)]}{F[F + 2\beta(F - 1)]}.
\end{equation}
In the case where $\beta = \gamma$, the present-day acceleration (\ref{present day acceleration}) reduces to 
\begin{equation}
\label{q for beta equal gamma}
\bar{q} = 2 - 3{\bar{\Omega}}_{m}/2.
\end{equation}
In particular, for $\Omega_{m} = 1$ we obtain $q = 1/2$ as in the dust FRW cosmologies.
We note that (\ref{q for beta equal gamma}) is positive for any physical value of $\Omega_{m}$, which means that a brane-universe with constant vacuum energy {\it must be slowing down} its expansion\footnote{This contradicts the observational fact that the universe is speeding up and not slowing down  (\ref{present q}). }. However, for $\beta \neq \gamma$, this is no longer so; the vacuum energy $\sigma$ is now a dynamical quantity which changes this picture. 

The choice of $\beta $ in (\ref{present day acceleration}) has to be consistent with recent measurements which indicate  that the current universe is speeding up its expansion with an acceleration parameter which is roughly 
\begin{equation}
\label{present q}
\bar{q} = - 0.5 \pm 0.2.
\end{equation}

\subsection{The cosmological constant}

Finally, for the cosmological parameter $\Lambda_{(4)}$ at  the present day we have
\begin{equation}
\label{cosmological parameter today}
{\bar{\Lambda}}_{(4)} = \xi(z_{eq}, \beta, {\bar{\Omega}}_{m}) {\bar{H}}^2,
\end{equation}
where 
\begin{equation}
\label{xi}
\xi(z_{eq}, \beta, {\bar{\Omega}}_{m}) = - \frac{3 \beta (F - 1)[{\bar{\Omega}}_{m}F^2 + \beta(F - 1)(1 + {\bar{\Omega}}_{m}F)]}{F[F + 2 \beta(F - 1)])}.
\end{equation}
Thus, ${\bar{\Lambda}}_{(4)} = 0$ for $\beta = \gamma$, as expected. Otherwise, for $\beta \neq \gamma = 0$, ${\bar{\Lambda}}_{(4)}$
has to be  positive in order to explain the present acceleration. 

\section{The auxiliary parameter $z_{eq}$}

 We now turn our attention to parameter $z_{eq}$. Firstly, we  confine the range of $z_{T}$, the redshift of transition from deceleration to acceleration, which is a solution of $q(z_{T}) = 0$. Secondly, we set a lower bound on the value of the deceleration parameter today. Thirdly, we study in some detail the models with ${\bar{\Omega}}_{m} = 0.3$. An interesting discovery here is the possibility of   
recollapsing models. In what follows  we will only  select $\beta$ from the range of values allowed by the  requirements (\ref{theoretical conditions on beta}), (\ref{upper bound for g})  and (\ref{present q}).

\subsection{Transition from deceleration to acceleration}

There is plenty of observational evidence for a decelerated universe in the recent past, see e.g. \cite{Riess2}-\cite{Turner3}.  
However,  the dominance  of the vacuum energy at some $z_{eq} > 0$ does not guarantee the present acceleration of the universe. For this, the vacuum energy has to dominate long enough as to overcome the gravitational attraction produced by ordinary matter. However, for every fixed value of ${\bar{\Omega}}_{m}$, we can find a   $z_{eq}^{min}$ for which the transition from deceleration to acceleration occurs at some $z_{T}$ ($0 < z_{T} < z_{eq}^{min}$). 

The transition is then guaranteed for  $z_{eq} > z_{eq}^{min}$. A straightforward numerical study of (\ref{acceleration}), in the interval $ (z_{eq}^{min}, \infty) $, reveals that $z_{T}$ is bounded above. Thus in our model, for every  ${\bar{\Omega}}_{m}$ the transition from deceleration to acceleration occurs in a finite redshift interval 
\begin{equation}
z_{T}^{min}  < z_{T} < z_{T}^{max},
\end{equation}
where the precise value of the lower and upper bounds depends on the density parameter for ordinary matter. The existence of an upper bound is an evidence that the dynamical influence of vacuum  energy rapidly decreases for redshifts $z > z_{T}^{max}$. We now proceed to show our numerical results for various values\footnote{A reliable and definitive determination of ${\bar{\Omega}}_{m}$ has thus far eluded cosmologists. 
However, the observational data   indicate that $\Omega_{\rho} \approx 0.1 - 0.3$ seem to be the most {\em probably} options.} 
 of ${\bar{\Omega}}_{m}$.

\paragraph{{$\bf{{\bar{\Omega}}_{m}} = 0.1$}:}

\medskip

We find that  a transition from deceleration to acceleration is only possible if the vacuum energy starts to dominate in an epoch before $z_{eq} > z_{eq}^{min} = 1.3$, otherwise the universe would be still today in a deceleration phase. The actual redshift of transition is $z_{T} = 1.1$. A detailed investigation of $q(z_{T}) = 0$, with $q$ given by (\ref{acceleration}), shows that $z_{T} < z_{T}^{max} = 1.81$ for $z_{eq}$ in the range $1.3 < z_{eq} < \infty$, i.e., the transition occurs in the interval $z_{T} = (1.10, 1.81)$.  
Thus, for the whole range of $z_{eq}$, we  find that  the redshift of transition from deceleration to acceleration, for ${\bar{\Omega}}_{m} = 0.1$, is\footnote{We note that these numbers are approximate; they depend on the specific choice of $\beta$ in the range that satisfies the conditions (\ref{theoretical conditions on beta}) and (\ref{upper bound for g}). However, since $|\Delta \beta/ \beta|$ is small and decreases with the increase of $z_{eq}$ (for an illustration see Table $1$),  the variation of these numbers is negligible  and does not change the picture here.}  
\begin{equation}
\label{Omega 0.1}
z_{T} \approx 1.46 \pm 0.36.
\end{equation}

\paragraph{ {$\bf{{\bar{\Omega}}_{m}} = 0.2$}:} The transition is guaranteed for $z_{eq} > 0.7$. It  occurs in the redshift interval $z_{T} = (0.58, 1.10)$, or equivalently 
\begin{equation}
\label{Omega 0.2}
z_{T} \approx 0.84 \pm 0.26,
\end{equation}
 and  {\it any} value of $z_{eq} > 0.7$.

\paragraph{{$\bf{{\bar{\Omega}}_{m}} = 0.3$}:} The vacuum contribution must start dominating at an epoch earlier than $z_{eq} = 0.4$, otherwise there would be no enough time for a transition from a decelerating phase to an accelerating one today. For any $z_{eq} > 0.4$, we find that the transition occurs at 
\begin{equation}
\label{Omega 0.3}
z_{T} \approx 0.48 \pm 0.20.
\end{equation}
This is consistent with the value $z_{T} \approx 0.5$ given by Turner and Riess \cite{Turner3}, and a little less than $z_{T} \approx 0.73$ provided by Perlmutter\footnote{It is encouraging   that completely different models provide similar values for the redshift of transition from deceleration to acceleration. See for example \cite{Bayin2}} 
 {\it et al} \cite{Perlmutter}.

The above discussion illustrates three things. Firstly, the fact that  the redshift $z_{eq}$, at which the vacuum energy starts to dominate,  depends on the mass density of ordinary matter;  in a low matter density universe the vacuum energy starts to dominate  before than in a universe with  high matter density. Secondly, that  the redshift of transition increases with the decrease of the matter density (\ref{Omega 0.1})-(\ref{Omega 0.3}). Thirdly, that the vacuum effects only become important at the present epoch, making the transition a recent phenomenon.

\subsection{Lower bound on present deceleration (Upper bound on acceleration)}

The deceleration parameter today, which is given by (\ref{present day acceleration}),  is an increasing  function of $\beta$, for any fixed values of  $z_{eq}$ and ${\bar{\Omega}}_{m}$. This means that its minimum value ${\bar{q}}_{min}$ is attained in the limit $\beta \rightarrow -1$. 

Now, fixing ${\bar{\Omega}}_{m}$ and selecting an appropriate $\beta$ (as to satisfy the conditions discussed in the previous section) we find that ${\bar{q}}_{min}$ decreases for large values of   $z_{eq}$. As an example consider the values presented in Table $1$; ${\bar{q}}_{min}$ increases   for $z_{eq} < 0.64$ and decreases  for $z_{eq} >  0.64$. Therefore,  the lowest value of ${\bar{q}}_{min}$ is attained in the limit $z_{eq} \rightarrow \infty$. Thus, from (\ref{present day acceleration}) we get 
\begin{equation}
\label{lower bound}
\bar{q} >  - 1 + \frac{3}{2}{\bar{\Omega}}_{m},
\end{equation}
for any value of $z_{eq}$. For  ${\bar{\Omega}}_{m} = 0.3$, we find $\bar{q} > - 0.55$, which is consistent with recent observations (\ref{present q}).

A similar analysis of (\ref{cosmological parameter today}) reveals that ${\bar{\Lambda}}_{(4)}$, the cosmological ``constant" today,  is bounded above. Namely, 
\begin{equation}
\label{Nothing new for Lambda}
{\bar{\Lambda}}_{(4)} < 3 {\bar{H}}^2 (1 - {\bar{\Omega}}_{m}),
\end{equation}
for any $z_{eq}$. Thus, if we take ${\bar{\Omega}}_{m} = 0.3$,  we get ${\bar{\Lambda}}_{(4)} < 2.1 {\bar{H}}^2$. However, this is just a statement that in the present model ${\bar{\Omega}}_{m} + {\bar{\Omega}}_{\Lambda_{(4)}} < 1$ in view of the quadratic correction in the generalized Friedmann equation.

\subsection{The model for ${\bar{\Omega}}_{m} = 0.3$}

Current dynamical mass measurements suggest that the matter content of the universe adds up to $30 \%$ of the critical density\footnote{Radiation $0.005\%$,  ordinary luminous baryonic matter $0.5 \%$, ordinary non-luminous baryonic matter $3.5 \%$ and exotic (non-baryonic) dark matter ``observed" through their gravitational effects $26 \%$.}. According to   (\ref{Omega 0.3})  
the transition from deceleration to acceleration occurs at a redshift $0.28 < z_{T} < 0.68$, which confirms the idea that the   accelerated expansion of the universe is a recent phenomenon.

Thus, in our model,  stars and galaxies with redshifts larger than $z \approx  0.68$ should reflect the kinematics of a decelerating  expansion. This is compatible with  galaxy formation, which can only take place if the gravitational attraction dominates a sufficiently long epoch over vacuum repulsion\footnote{For an updated analysis of the influence of dark energy on the first epoch of galaxy formation see, e.g., \cite{Lima}}. It also fits the  observations of SN $1997$ff at $z \approx 1.7$. This is the oldest and most distant type Ia supernova (SN Ia) discovered so far and provides direct evidence that the universe was decelerating before it began speeding up  \cite{SN1977ff}. Other high redshift SNe known at $z = 1.2$ (e.g., SN $1999$fv, SN $1998$eq), $z = 1$ (e.g., SN $1997$ck, SN $1999$fk) and $z = 0.83$ (e.g., SN $1996$cl) can provide a direct test  for  deceleration at the time of their explosion.

Also, from (\ref{Omega 0.3}) it follows that galaxies with redshifts less that  $z \approx 0.28 $ should show evidence of an accelerating universe. Very-low redshift supernovae are crucial for reducing the uncertainty of the contemporary expansion rate.

We now proceed to study in more detail the parameters of the model for  ${\bar{\Omega}}_{m} = 0.3$, which is favored by observations. First of all, let us simplify the notation. To this end, we introduce the dimensionless parameters $\eta_{k}$ and $\eta_{\sigma_{0}}$ as
\begin{equation}
\label{new notation}
\eta_{k} \equiv \frac{k_{(5)}^4{\bar{\rho}}_{m}^2}{18{\bar{H}}^2}, \;\;\;\eta_{\sigma_{0}} \equiv \frac{\sigma_{0}}{{\bar{\rho}}_{m}},
\end{equation}
in terms of which the generalized Friedmann equation (\ref{F. eq. for the case under study}) becomes
\begin{equation}
\label{Friedmann eq. in new notation}
{H}^2 = {\bar{H}}^2{\eta_{k}}\left[\eta_{\sigma_{0}} + \frac{1}{2}\left(\frac{\rho_{eff}}{{\bar{\rho}}_{m}}\right)\right]\left(\frac{\rho_{eff}}{{\bar{\rho}}_{m}}\right),
\end{equation}
We note that 
\begin{equation}
\eta_{k}\eta_{\sigma_{0}}  F + \frac{\eta_{k}}{2}F^2 = 1,
\end{equation}
which follows from (\ref{sigma 0 as a function of the parameters and rho bar}) and (\ref{k}), so that $H = \bar{H}$ today, as expected.

Examination  of (\ref{total sigma}), (\ref{k}) and (\ref{g}) reveals that, for every given $z_{eq}$ and ${\bar{\Omega}}_{m}$, the adequate values of $\beta $, that satisfy the conditions (\ref{theoretical conditions on beta}) and (\ref{upper bound for g}), are spread over a small range. In that range $\bar{q}$,  $\eta_{k}$ and $\eta_{\sigma_{0}}$ increase with $\beta$, while $\bar{\sigma}$ and ${\bar{\Lambda}}_{(4)}$ decrease with $\beta$. For example, if we take $z_{eq} = 1$ and ${\bar{\Omega}}_{m} = 0.3$, we find\footnote{For $z_{eq} = 1$ and ${\bar{\Omega}}_{m} = 0.3$,  any $\beta$ in the interval $(-1, -0. 65)$ satisfies $k_{(5)}^4 > 0$ (or $\eta_{k} > 0$) and $\bar{\sigma} > 0$. However, the condition $|g| < 0.1$ narrows down this interval to  $(-1, - 0.978)$.} $\beta = - 0.989 \pm 0.011$. The corresponding $\bar{q}$ increases from $\bar{q} = - 0.46$ for $\beta \approx - 1$ to $\bar{q} = -0.35$ for $\beta = - 0.978$. The other quantities undergo a relatively smaller change, viz.,
\begin{eqnarray}
\label{values of the cosmological parameters for Zeq = 1and Omega matter 0.3}
\eta_{k} = 0.056^{+ 0.002}_{- 0.002}, \;\;\; \eta_{\sigma_{0}} = - 2.390^{+ 0.049}_{- 0.046},\;\;\;\frac{\bar{\sigma}}{{\bar{\rho}}_{m}} = 5.343^{+ 0.196}_{- 0.210},;\;\;\; \frac{{\bar{\Lambda}}_{(4)}}{{\bar{H}}^2} = 1.923^{+ 0.087}_{- 0.094}.
\end{eqnarray}

For the redshift of transition from deceleration to acceleration we find
\begin{equation}
z_{T} = (0.335, 0.298, 0.279) \;\;\; \mbox{for}\;\;\; \beta = (- 0.999, - 0.985, - 0.978)\;\;\;   \mbox{respectively}.
\end{equation}

In Table 1, we illustrate the cosmological parameters for various values of $z_{eq}$ and ${\bar{\Omega}}_{m} = 0.3$.  For the sake of simplicity,  we omit their small change over the range of $\beta$ and  present  their mean values only\footnote{See footnote 9}, corresponding to the average of $\beta$. Like we have said before the vacuum contribution must start dominating at an epoch earlier than $z_{eq} = 0.4$, otherwise the universe would be still in a decelerating phase. 

\begin{center}
\begin{tabular}{|c|c|c|c|c|c|c|c|c|} \hline
\multicolumn{9}{|c|}{\bf Table 1: Cosmological parameters for ${\bar{\Omega}}_{m} = 0.3$ fitting $|g| <  0.1$}\\ \hline
 \multicolumn {1}{|c|}{$z_{eq}$} & $- 1 < \beta <$&$\eta_{k}$& $(\bar{\sigma}/{\bar{\rho}}_{m})$ & 
$\bar{q}$ & $\eta_{\sigma_{0}}$&$\bar{H}\bar{t}$&$({\bar{\Lambda}}_{(4)}/{\bar{H}}^2)$& 
 \multicolumn{1}{|c|}{$z_{T}= z_{q = 0})$} \\ \hline\hline
$0.4$ &$- 0.840$ & $0.030$  & $10.052$ &$(- 0.49, - 0.38)$&$7.724$&$0.743$&$1.853$&$0.378$\\ \hline
$0.5$ &$ - 0.949$ & $0.063$  & $4.791$ &$(- 0.46, - 0.35)$&$1.602$&$0.691$&$1.915$&$0.293$ \\ \hline
$0.6$ &$- 0.963$ & $0.070$   & $4.281$& $(- 0.45, - 0.34)$&$0.364$&$0.681$&$1.912$&$0.277$ \\ \hline
$0.7$ &$- 0.970$ & $0.070$  & $4.298$ &$(- 0.45, - 0.34)$&$- 0.427$&$0.682$&$1.915$&$0.278$  \\ \hline
$0.8$ &$- 0.974$ & $0.066$  & $4.529$ &$(- 0.45, - 0.35)$&$- 1.097$&$0.686$&$1.919$&$0.286$  \\ \hline
$0.9$ &$- 0.976$ & $0.061$  & $4.885$ &$(- 0.46, - 0.35)$&$- 1.737$&$0.693$&$1.920$&$0.297$ \\ \hline
$1.0$ &$- 0.978$ & $0.056$  & $5.343$ &$(- 0.46, - 0.35)$&$- 2.390$&$0.701$&$1.923$&$0.311$ \\ \hline
$1.1$ &$- 0.979$ & $0.051$  & $5.879$ &$(- 0.47, - 0.35)$&$- 3.073$&$0.708$&$1.923$&$0.324$ \\ \hline
$1.2$ &$- 0.979$ & $0.046$  & $6.482$ &$(- 0.48, - 0.35)$&$- 3.796$&$0.716$&$1.916$&$0.336$  \\ \hline
$1.5$ &$- 0.981$ & $0.034$  & $8.768$ &$(- 0.49, - 0.34)$&$- 6.309$&$0.738$&$1.903$&$0.375$  \\ \hline
$1.7$ &$- 0.982$ & $0.028$  & $10.673$ &$(- 0.50, - 0.32)$&$- 8.317$&$0.751$&$1.887$&$0.398$  \\ \hline

 \end{tabular}
\end{center}

The second column in Table $1$ gives the range of $\beta$, for different values of $z_{eq}$, that satisfy the requirements $k_{(5)}^4 > 0$, $G > 0$ and $|g| < 0.1$. This is illustrated in columns $3$ and $4$ by the positive values of $\eta_{k}$ and $(\bar{\sigma}/{\bar{\rho}}_{m})$, respectively. 

For these values, it turns out that the acceleration of the universe today is {\it not} an additional   constraint or assumption. On the contrary,  it is a consequence of  the model and the observational requirement $|g| < 0.1$. The significant point here is that the  results, which are presented in column $5$, are consistent with current observations (\ref{present q}). Besides, our model narrows down the possible values of $q$ today (\ref{lower bound}).

For larger values of $z_{eq}$, say $z_{eq} > 1.8$, the requirements (\ref{theoretical conditions on beta}) and (\ref{upper bound for g}) are automatically satisfied if $\beta$ is selected in the range $(- 1, - 0.983)$. However, the cosmic  acceleration (\ref{present q}) is not a consequence of the model (as it is for $z_{eq} \leq 1.8$) but an observational constraint which restricts the values of $\beta$ even further.

Column $6$ presents the dimensionless parameter $\eta_{\sigma_{0}}$. It changes sign for $z_{eq} \approx 0.64$. Therefore, there are two kinds of models:

(i) {\it Ever-expanding models}. These are the  ones with $z_{eq} < 0.64$, for which $\eta_{\sigma_{0}} > 0$. After a long matter-dominated phase of deceleration, the universe is accelerating today and will continue that way in the future\footnote{In order to avoid misunderstanding, let us mention that the equation $\sigma_{0}(z_{eq}, \beta) = 0$ does have a solution for $z_{eq}  = (0.4, 0.5, 0.6, 0.64)$ and $\beta = (- 0.67, - 0.72, - 0.83, - 0.96)$, respectively. But these values of $\beta$ do not satisfy the physical conditions (\ref{theoretical conditions on beta}), (\ref{upper bound for g})  and (\ref{present q}). Therefore they are excluded here. More precisely, the only models with $\sigma_{0} = 0$ that satisfy physical conditions are those with $z_{eq}$ in the interval $(0.642, 0.643)$ and $-1 < \beta < - 0.967$.}.

(ii) {\it Recollapsing models}. For $z_{eq} >  0.64$, we have $\eta_{\sigma_{0}} < 0$.  After a long matter-dominated phase of deceleration, the universe is accelerating today, but deceleration comes back at some point in the future\footnote{For $0.64 < z_{eq} < \infty$,  using (\ref{acceleration}) we find that equation $q(z_{T}) = 0$ has two set of solutions. One for positive $z$,  which is  given by (\ref{Omega 0.3}), and another one for negative $z$, namely  $ - 1 <z_{T} < - 0.22$. The second solution corresponds to the future,  when the size of the universe is $a  > 1.28 \bar{a}$.}.  From (\ref{Friedmann eq. in new notation}) it follows that the expansion must come to an end before the universe starts contracting. Setting $H = 0$ we can get $a_{rec}$, the size of the universe at the moment of recollapse. Namely, 
\begin{equation}
\label{size at recollapse}
\left(\frac{\bar{a}}{a_{rec}}\right)^3 + (F - 1)\left(\frac{\bar{a}}{a_{rec}}\right)^{3(\beta + 1)} + 2\eta_{\sigma_{0}} = 0.
\end{equation}
Clearly, in the case under consideration the second term amply dominates over the first one. Therefore, from (\ref{size at recollapse}) we get
\begin{equation}
\label{value of a at rec}
a_{rec}\approx \bar{a}\left(\frac{F - 1}{2|\sigma_{0}|}\right)^{1/{3(\beta + 1)}}, \;\;\;\beta \neq - 1.
\end{equation}
The reason for the recollapse is that the four dimensional cosmological constant changes its sign sometime in the far future. Using (\ref{cosmological constant in terms of a}) and (\ref{value of a at rec}), we find 
\begin{equation}
\label{size for the vanishing of the cosm constant} 
a_{\Lambda_{(4)} = 0} \approx |\beta|^{1/{3(\beta + 1)}}a_{rec}.
\end{equation}   
As an example, we use  the data for $z_{eq} = 0.7$. We find
\begin{equation}
 a_{rec} \approx 4.44 \times 10^{16} \bar{a}, \;\;\;a_{\Lambda_{(4)}} \approx 0.71 a_{rec}.
\end{equation}
The good news is that  we are nowhere near the recollapse! 

The age of the universe $\bar{t}$,  in terms of the current value of the Hubble ``constant" $\bar{H}$,  is presented in the seventh column. The main feature here is that the universe is older than the FRW counterpart for which $\bar{H}\bar{t} = 2/3$. To put the discussion in perspective,  we notice that in braneworld models (with $\Lambda_{(4)} = 0$) the universe  is {\it younger}  than in the standard FRW cosmologies \footnote{For example,  for a dust-filled universe without  cosmological term, we find ${\bar{t}}_{brane} = 0.363{\bar{H}}^{- 1}$ in brane models, while ${\bar{t}}_{FRW} = 2{\bar{H}}^{- 1}/3$ in FRW models.}. The age of the universe increases with $z_{eq}$. As an example, if we take  $z_{eq} = 100$ (for this value the crossover point is $z_{T} = 0.67$),  we find $\bar{H}\bar{t} = 0.96$.

Finally, the redshift of transition is specified in column $9$.  For the values considered here the crossover  takes place  at $z_{T} \sim 0.28 - 0.4$. However, it can occur  even earlier  
if the vacuum contribution starts to dominate for larger values of $z_{eq}$, in which case the range for $z_{T}$ extends to $0.28 < z_{T} < 0.68$.
 This redshift  
interval coincides with the time of explosions of a number of SNe Ia known today. For example, SN $1988$ at $z = 0.31$, SN $1992$bi at $z = 0.46$, SN $1995$K at $z = 0.48$ and 
SN $1995$  at $z = 0.66$. Therefore, they can provide crucial information to reduce the uncertainty in the transition  between deceleration and acceleration.

For completeness, we  mention that the cosmological constant in the bulk, using  (\ref{avoiding exp. and rec.}) and (\ref{new notation}), can be written as 
\begin{equation}
\Lambda_{(5)} = - \frac{\eta_{\sigma_{0}}^2\sqrt{\eta_{k}}}{\sqrt{2}}\bar{H}{\bar{\rho}}_{m}.
\end{equation}
Thus, we can ``predict" the value of $\Lambda_{(5)}$ by means of  measurements in $4D$.

\section{Discussion and conclusions}

An important distinction between general relativity and braneworld theory is that the cosmological term in the first is put by ``hand" while in the second  it is determined  by the  solution in the bulk through Israel's boundary conditions. 

A variable vacuum energy can generally occur as a consequence of embedding our universe as a brane in a five-dimensional bulk with non-static extra dimensions.   There are a number of known solutions in $5D$ for which the vacuum  density on the brane decreases as an inverse power of the scale factor, similar (but at different rate) to the power law in FRW-universes of general relativity.

In this paper, we devoted our attention  to spatially flat, homogeneous and isotropic, brane-universes where the vacuum  density decays as in (\ref{variable density}). The model contains two parameters, viz., $\beta$ and $z_{eq}$. 

If $\beta = \gamma$, then $\Lambda_{(4)} = 0$, $G$ is a universal constant and  $\bar{q} = 2 - 3{\bar{\Omega}}_{m}/2$,  in the dust dominated era. Thus fixing $\Omega_{m}$ today also fixes $q$. Notice that $q$ is positive for any physical value of $\Omega_{m}$, meaning that a brane-universe with constant vacuum energy must be slowing down its expansion. 

However, for $\beta \neq \gamma$, this is no longer so; the vacuum energy is now a dynamical quantity which changes this picture. In fact, G as well as $\Lambda_{(4)}$ become functions of time and the deceleration parameter decreases from $q_{\gamma} \approx 3\gamma + 2$ at the beginning of the universe to $q \rightarrow - 1 + 3\Omega_{m}/2$ at the present time. Thus, for any matter with $\gamma > - 2/3$, $q$ is positive at the beginning  and negative today, because currently ${\bar{\Omega}}_{m} < 2/3$.

Our model predicts that the transition from deceleration to acceleration  occurred only recently, for $z_{T} < 0.68$, but not later than $z_{T} \approx 0.28$, regardless of the specific value of $z_{eq}$. Therefore, early structure formation, from small density inhomogeneities, is not affected.

If the domination of the vacuum is recent, i.e.,  $z_{eq} < 1.8$, then the observed accelerated cosmic expansion is,  {\em not} a condition but, a consequence of our model. The truth is that we have no observational or theoretical reasons to believe that vacuum started to dominate before $z \sim 1$.  
Consequently, we can safely declare that our model predicts the present acceleration of the universe. 

What is important here is that  the predicted values for $\bar{q}$ are consistent with observations and allow us to narrow down the experimental uncertainty in the current data; from (\ref{present q}) to

\begin{equation}
\label{using the model to narrow down the interval for q}
\bar{q} = - 0.41 \pm 0.09. 
\end{equation}
This may help in observations for an experimental/observational test of the model.

Let us notice that an alternate way to write the effective density (\ref{eff. density in terms of F}) is 

\begin{equation}
\label{contact with cardasian models}
\rho_{eff} = \rho_{m} + \frac{(F - 1)}{{\bar{\rho}}_{m}^{(n - 1)}}\rho_{m}^n, 
\end{equation}
where $n = (\beta + 1)$.
This peculiar notation allows us to make contact between our ever-expanding models (those with $\sigma_{0} > 0$) and  Cardassian models. Indeed, for $\sigma_{0} \gg \rho_{eff}/2$ the second term  in (\ref{F. eq. for the case under study}) can be neglected\footnote{This is clearly satisfied for $z_{eq} \approx 0.4$ during the recent epoch of accelerated expansion which started at $z \approx 0.38$. See Table $1$}. In this approximation, the generalized Friedmann equation (\ref{F. eq. for the case under study}) can be written as
\begin{equation}
\label{Cardassian models}
H^2 \approx A \rho_{m} + B \rho_{m}^{n}, 
\end{equation}
with $n = (\beta + 1)$ and
\begin{equation}
\label{coeff. for the Cardassian model}
A = \frac{{\bar{H}}^2\eta_{k}\eta_{\sigma_{0}}}{{\bar{\rho}}_{m}}, \;\;\;B = \frac{{\bar{H}}^2\eta_{k}\eta_{\sigma_{0}}(F - 1)}{{\bar{\rho}}_{m}^n}.
\end{equation}
Equation (\ref{Cardassian models}) is similar to the one used in the so-called Cardassian models \cite{Freese}. It differs from the usual Friedmann equation of general relativity by the addition of the extra term $\rho_{m}^{n}$.  Therefore, based on (\ref{Cardassian models}),  one can interpret Cardassian expansion as the  low energy limit of the brane model discussed here.  Freese and Lewis \cite{Freese} and Gondolo and  Freese \cite{Gondolo} suggest that the extra term may arise as a consequence of embedding our universe as a brane in extra dimensions. This is exactly what we have here. 

Nonetheless, there is a big difference in the understanding of the extra term. In the Cardassian model, by assumption,  there is no vacuum contribution and the new term may come from some (yet unknown) modified Einstein equations. In our model the extra term is a manifestation of the variation of vacuum, in response to the time evolution of the extra dimension

This difference leads to distinct requirements on $n$ (or $\beta$). In Cardassian models the most stringent requirements on $n$ demand $n < 0.4$ (equivalently, $\beta < -  0.6$). 
While, in our model the parameter $\beta$  depends on ${\bar{\Omega}}_{m}$ and is severely restricted by physical conditions $k_{(5)}^4 > 0$, $G > 0$ and $|g| < 0.1$. Indeed, for ${\bar{\Omega}}_{m} = 0.3$ we found $-1 < \beta < - 0.84$ (equivalently, $ 0 < n < 0.16$). 

In order to get another  interpretation  we write (\ref{Cardassian models}) as
\begin{equation}
\label{quintessence}
H^2 \approx {\bar{H}}^2\left[\Omega_{M}(1 + z)^3 + \Omega_{X}(1 + z)^{3(w_{X} + 1)}\right],
\end{equation}
with $w_{X} = \beta$, $\Omega_{M} = \eta_{k}\eta_{\sigma_{0}}$ and $\Omega_{X} = \eta_{k}\eta_{\sigma_{0}}(F - 1)$. In this approximation $F \approx (\eta_{k}\eta_{\sigma_{0}})^{- 1}$, so that $\Omega_{M} + \Omega_{X} \approx 1$. The above expression  is simikar to the Friedmann equation in cosmological models with quintessence, where the dark energy component is characterized by the equation of state $w_{X} = p_{X}/\rho_{X}$. 

Thus, if the vacuum energy only started to dominate recently, at $z_{eq} \sim 0.4 - 0.6$, then at low energy limit there is a correspondence between our ever-expanding as well as  the Cardassian and  quintessence models, although they are entirely different in philosophy.

However, if the vacuum energy started to dominate earlier, say at $z_{eq} > 0.64$, then $\sigma_{0}$ is negative and the quadratic term in  (\ref{F. eq. for the case under study}) cannot be disregarded. As a consequence our model (for $z_{eq} > 0.64$) is entirely different from Cardassian and quintessence models. In particular, it predicts that our universe will collapse in the future.

 We would like to finish this work with the following comments. 
The whole analysis  in this paper is 
independent of any particular solution used in the five-dimensional bulk. 
This is a great virtue of brane-world models as noted at the end of Section $2$. 
However, one could still ask whether the universe described here 
can be embedded in a five-dimensional bulk. The answer to this question 
is positive. For example, they can be embedded in 
five-dimensional ``wave-like" cosmologies of the type 
discussed in \cite{jpdel4}. If in equation $(38)$ of \cite{jpdel4} 
we take variable $\sigma$ as here in (\ref{variable density}), then the scale 
factor $a$ for such wave-like models is governed by an equation 
which is identical to (\ref{gen. F. eq. for the case under study}) 
in this paper.

It is important to mention that the ratio $(\dot{\Phi}/\Phi)$ appears  in different contexts, notably  in expressions concerning the variation of rest mass \cite{defofmass}- \cite{Extra force},  electric charge \cite{massandcharge} and variation of the gravitational ``constant" $G$ \cite{Melnikov3},\cite{Melnikov1}.  Therefore, we have a scenario where  the observed cosmic acceleration is just one piece in the dynamical evolution of an universe where the so-called fundamental ``constants" are evolving in time. Braneworld scenario may provide us a theoretical framework  to unify all these, apparently, separated  phenomena as different consequences in $4D$ of the  time evolution of the extra dimension. This is a new step toward  understanding how the universe works.

{\bf Note added in proof:} While different alternative explanations are given to explain the current acceleration of the universe \cite{Caldwell2}-\cite{Vish and Singh}, the dimming of the supernovae could be explained on the basis of axion physics \cite{Csaba Csasi et al}.

\end{document}